\documentstyle[12pt,epsf,epsfig,amstext]{article} 
\chardef\letterchar=11
\chardef\otherchar=12
\chardef\eolinechar=5
\ifx\selectfont\undefined%
  \def\mathrm#1{{\rm #1}}
\fi
\def\ifmath#1{\relax\ifmmode #1\else $#1$\fi}%
\let\ifmathx0
\def\fixmath{\def\ifmath{\noexpand\ifmathx}}%

\def\ovMZ{\ifmath{{\overline M}_Z}}

\def\ovMZ{\ifmath{{\overline m}_Z}}
\def\ovGZ{\ifmath{{\overline \Gamma}_Z}}

\def\gam{\ifmath{\gamma}}%
\def\Zo{\ifmath{\mathrm {Z}}}

\def\Rgf{\ifmath{\mathrm {R}^f_\gam}}

\def\RZfi{\ifmath{\mathrm {R_Z}^{fi}}}

\def\Ffin{\ifmath{\mathrm {F}_n^{fi}}}

\def\RZfn{\ifmath{\mathrm {R_Z}^{f0}}}
\def\RZfe{\ifmath{\mathrm {R_Z}^{f1}}}
\def\RZfz{\ifmath{\mathrm {R_Z}^{f2}}}
\def\RZfd{\ifmath{\mathrm {R_Z}^{f3}}}

\newcommand{\imag}{\mbox{$\Im$m}}
\newcommand{\real}{\mbox{$\Re$e}}

\newcommand{\noi}{\noindent}

\newcommand{\bq}{\begin{equation}} 
\newcommand{\eq}{\end{equation}}
\newcommand{\ba}{\begin{eqnarray}}
\newcommand{\ea}{\end{eqnarray}}

\newcommand{\nl}{\\ \nonumber}
\begin{document}
  \psfull
\begin{flushleft}
DESY 97--001
\\
hep-ph/9709208
~v.2
\\
August 1997
\end{flushleft}
\noindent
\vspace*{0.50cm}
\begin{center}
\vspace*{2.cm} 
{\huge  The $Z$ Boson Resonance Parameters
    \vspace*{1.cm}
}
\bigskip
\\ {\Large
Tord Riemann}
\bigskip
\\  {\large
DESY -- Institut f\"ur Hochenergiephysik
    \vspace*{0.3cm}
\\
Platanenallee 6, D-15738 Zeuthen, Germany}
\\
\bigskip
{\small
Email: riemann@ifh.de \,\, $\cdot$ \,\,
http://www.ifh.de/\,$\tilde{}$\,riemann
}
\bigskip
\vspace*{1.cm}

\noi 
{\em 
Talk held at the Symposium on Semigroups and Resonances 
\\
at the XXI
Int. Colloquium on Group Theoretical Methods in Physics
\\ 15--20 July
1996, Goslar, Germany
}
\bigskip
\end{center}

\vfill
 
\centerline{ABSTRACT}

\bigskip

\noi
The Z line shape is measured at LEP/CERN with an accuracy at the
  per mille level.
Usually it is described in the Standard Model of electroweak
  interactions with account of quantum corrections.
Alternatively, one may attempt an S-matrix based
  model-independent approach in order to extract quantities like mass
  and width of the Z boson.
I describe the formalism and its application to data.

\vfill
\bigskip
\vspace*{1.cm}

\newpage
\section{Introduction}
Our present understanding of weak interactions is completely described by the 
Standard Model~\cite{SM}, a spontaneously broken locally gauge invariant,
anomaly-free, renormalizable 
quantum field theory of pointlike leptons and quarks, the latter in three
colors. 
The model contains fermions, vector bosons, and a scalar particle:
\bigskip

fermions:
\begin{minipage}[t]{9.cm}     
                              \begin{minipage}[h]{2.cm}     
                                \[ \left( \begin{array}[h]{c} \nu_l \\ 
                                l
                               \end{array}
                             \right)
                             \]
                             \end{minipage}
                             ~~~~~~~$l=e,\mu,\tau$ \\ 
                              \begin{minipage}[h]{2.cm}     
                                \[ \left( \begin{array}[h]{c} U \\ D
                               \end{array}
                             \right)
                             \]
                             \end{minipage}
                              \begin{minipage}[h]{6.cm}     
                                \[ \left.
                              \begin{array}{l}
                                U=u,c,t \\ D=d,s,b
                              \end{array} \right\}   \mbox{in 3 colors}
                              \]
                              \end{minipage} 

\end{minipage}


\bigskip

\noi
vector gauge bosons: ~~~~~~~~~~~~$W^{\pm},~Z^0,~\gamma$

\bigskip

\noi
scalar Higgs boson: ~~~~~~~~~~~~~ $H$

\vspace*{1.cm}

The particle's masses (and mixing angles) are free parameters. Their
interactions are 
determined from the invariance of the Lagrangean under local gauge
transformations with gauge group $SU(2)_L\times U(1)$ and associated gauge
fields $W^{\pm}, W^0, B$.
One may parameterize the model in terms of masses and mixing angles plus
electromagnetic coupling constant $\alpha_{em}$.
Often, instead of the $W$ boson mass, the Fermi constant is used:
\begin{itemize}
\item $\alpha_{em} = 1/137.036$
\item $G_{\mu} = 1.16634 \times 10^{-5}$ GeV$^{-2}$
\item $m_{f}$, including $m_t = 175$ GeV~\cite{blondel}
\item $M_Z = 91.186$ GeV~\cite{blondel}
\item $M_H$ \ldots unknown
\end{itemize}

\bigskip

Discovery and study of the $Z$ resonance are part of the long history of weak
interactions and of unification of forces.
First observations of virtual $Z$ exchange lead to the discovery of weak 
neutral current reactions in the scattering of neutrinos off electrons,
$\nu_{\mu} + e^- \to   \nu_{\mu} + e^-$,  and off nucleons,  $\nu_{\mu} + N \to
\nu_{\mu} + N $ in 1973 (Gargamelle Collab.: F.J.~Hasert et al., A.~Benvenuti
et al., B.~Aubert et al.) at the proton accelerator PS (CERN). 
The cross-section measurements may be interpreted in terms of the weak mixing
angle $\theta_w$. 
This angle characterizes not only the mixing of photon and $Z$ boson
but also the strength of the weak neutral interactions and the relations of
the gauge boson masses to the Fermi constant (and among themselves).
In the Standard Model\footnote{
The relations get modified by radiative corrections; see e.g.~\cite{analysis}.
}:
\ba
Z &=& \cos\theta_w \, W^0 - \sin\theta_w \, B
\\
\gamma &=& \sin\theta_w \, W^0 + \cos\theta_w \, B
\\
g~{\sin \theta_w} &=& e = \sqrt{4\pi\alpha_{em}}
\\
a_{lept} &=& -\frac{1}{2}
\\
v_{lept} &=& -\frac{1}{2}\left(1-4\sin^2\theta_w\right) 
\\
M_W &=& \sqrt{\frac{\pi\alpha}{G_{\mu}\sqrt{2}}}~\frac{1}{\sin\theta_w} \geq
37.281 \, \,{\rm GeV} 
\\
M_Z &=& \frac{M_W}{\cos \theta_w}
\ea
The theory predicts the gauge boson masses as soon as
there is a numerical estimate for the weak mixing angle\footnote{
An absolute lower limit is about 37 GeV, see (6).
}.
Thus, after a few weak neutral current events were observed, a lot of
information could gained from this.
From the cross-sections of 1973, one may derive $0.1 < \sin^2 \theta_w < 0.6$. 
This corresponds to $M_W = 118 \cdots 48$ GeV and $M_Z =125 \cdots 75$ GeV.
Both particles were discovered at the specially designed $p{\bar p}$ collider
SPS (CERN) in 1983 
(UA1 Collab.: G.~Arnison et al., UA2 Collab.: P.~Bagnaia et al., M.~Banner et
al.). 

\bigskip

After the discovery of the $Z$ boson,
its detailed study by a dedicated tool, an $e^+e^-$ collider
with a center of mass energy corresponding to the $Z$ mass, became a dream of
particle physicists.
Since the advent of the $e^+e^-$ colliders LEP (CERN) and
SLC (SLAC) in 1989, about sixteen millions of $Z$ bosons have been produced at
LEP and hundreds of thousands at SLC. 
They are produced as a resonance peak in the cross-section of the reaction 
\ba
e^+ + e^- \quad  \to (\gamma, Z) \to \quad \text{anything}  
\label{1}
\ea
LEP finished operation as a $Z$ factory in 1995 and is now running at
higher energies for the study of $W$ pair production and searches
for Higgs, susy, and other particles while SLC goes yet on for a while.

Due to the impressive accuracy of the measurements it was possible to test
the Standard Model at the level of quantum corrections. 
This raises the problem of the accurate description of unstable
particles in a quantum field theory.
In a quantitative sense, this has been done with great success.
Practically all experimental results are described by the Standard Model
consistently within the experimental errors. 
I should mention specially the recent discovery of the top quark with a mass of
about 176 GeV at Fermilab (CDF Collab.: F. Abe et al. (1994), D0 Collab.: S.
Abachi et al. (1995)). 
This value agrees nicely with that predicted from measurements of the $Z$
resonance parameters when quantum corrections from virtual top
quark exchange are taken into account in the Standard Model, $m_t \sim
147 - 167$ GeV~\cite{PPE/96-183}. 
The fits favor a light Higgs boson with $M_H = 121_{-68}^{119}$ GeV and the
estimate $M_H < 430$ GeV at 95\% C.L. 
However, one should note
that there is no experimental hint for the existence of the Higgs boson, whose
interactions are assumed to create all the particle masses.

\bigskip

In this contribution, the shape of the $Z$ resonance excitation
curve will be described\footnote{
For the description of hadron resonances
see the contribution of G. L\'opez Castro~\cite{lopez} to this symposium.
}. 
It is analyzed to which extent the description is model-independent.
Some emphasis will be given to an approach based on first principles as
formulated in the S-matrix theory.

The $Z$ resonance is part of our physical world. 
We are not faced with the problem of its existence but rather of its
proper description.
A rigid mathematical handling of unstable particles 
to which the efforts of
many of the participants at this Symposium are devoted is certainly not
developed within the framework of relativistic quantum field theory.
I hope that my talk may serve as an introduction to the status of the study of
the $Z$ resonance. 
The presentation will reflect my working activities which are closely related
to the interpretation of measured cross-sections and other observables in terms
of theoretical quantities.
This task  
deserves a close interaction of theoreticians and experimental physicists.
For details about this cooperation interested collegues may consult
e.g.~\cite{blondel,analysis} and references therein. 

With the advent of the high-precision data from LEP~1 on single $Z$ boson
production and the frequent $W$ pair production at LEP~2, the problem
of definition of their masses and widths in a renormalizable quantum
field theory became an important issue.
Experimentalists often use formulae in the on-mass-shell approach, while
some theorists prefer the introduction of a complex particle pole
prescription, proposed in~\cite{consoli,stuart262}.
The first one is preferred by recent tradition and well-developed
while the latter one looks more convincing from a conceptual point of view:
the propagator may be constructed in an explicitely gauge invariant way.
When used properly, both schemes will give gauge-invariant results in
the relevant order of perturbation theory (see
e.g.~\cite{consoli,jegerTasi,sirlin,argyres}),
but the numerical values for the $Z$ mass differ significantly.
This was observed first in~\cite{BBHvN,leike}.  
Quite recently, the relation of both schemes was discussed in
detail~\cite{beenakker}. 
Although I will not give an introduction to perturbative renormalization for
unstable particles, few comments on it may be found in sections~\ref{I}
and~\ref{comm1}.  
\section{The $Z$ Line Shape
}
Some of the predictions of the Standard Model have been mentioned in the
Introduction. 
Particle masses are used as input parameters while their life times $\tau = 1 /
\Gamma, \Gamma$ being the decay width, may be predicted.
The $Z$ boson decays nearly exclusively into pairs 
$f {\bar f}$ of
leptons ($e, \nu_e, \mu, \nu_{\mu}, \tau, \nu_{\tau}$) or colored quarks (3
$\times$ $d, u, s, c, b$). 
The inverse life time (total $Z$ width) is the incoherent sum of all partial
widths of the different decay channels. 
From the Lagrangean for the $Zf{\bar f}$ interactions
\ba
{\cal L}  
&=& 
-\frac{ig}{2\cos\theta_w} \, 
Z_{\mu} \sum_f {\bar f} \gamma^{\mu} \left[ v_f +
a_f \gamma_5 \right] f  
\ea
one may derive~\cite{akhund1,analysis} 
\ba
\Gamma_Z &=& \sum_f \Gamma_f =
\sum_f N_{color}^f \frac{G_{\mu}M_Z^3}{6\pi\sqrt{2}}\left[
(v_f^{eff})^2+(a_f^{eff})^2\right]  =
2.4946 \pm 0.0027 ~\mbox{GeV}~\cite{blondel}
\nl
\ea
The notations indicate that the (effective) couplings are slightly modified by
radiative corrections.  

\begin{figure}[thbp]
 \vspace*{-6cm}
 \hspace*{-20mm}
\mbox{
        \epsfig{file=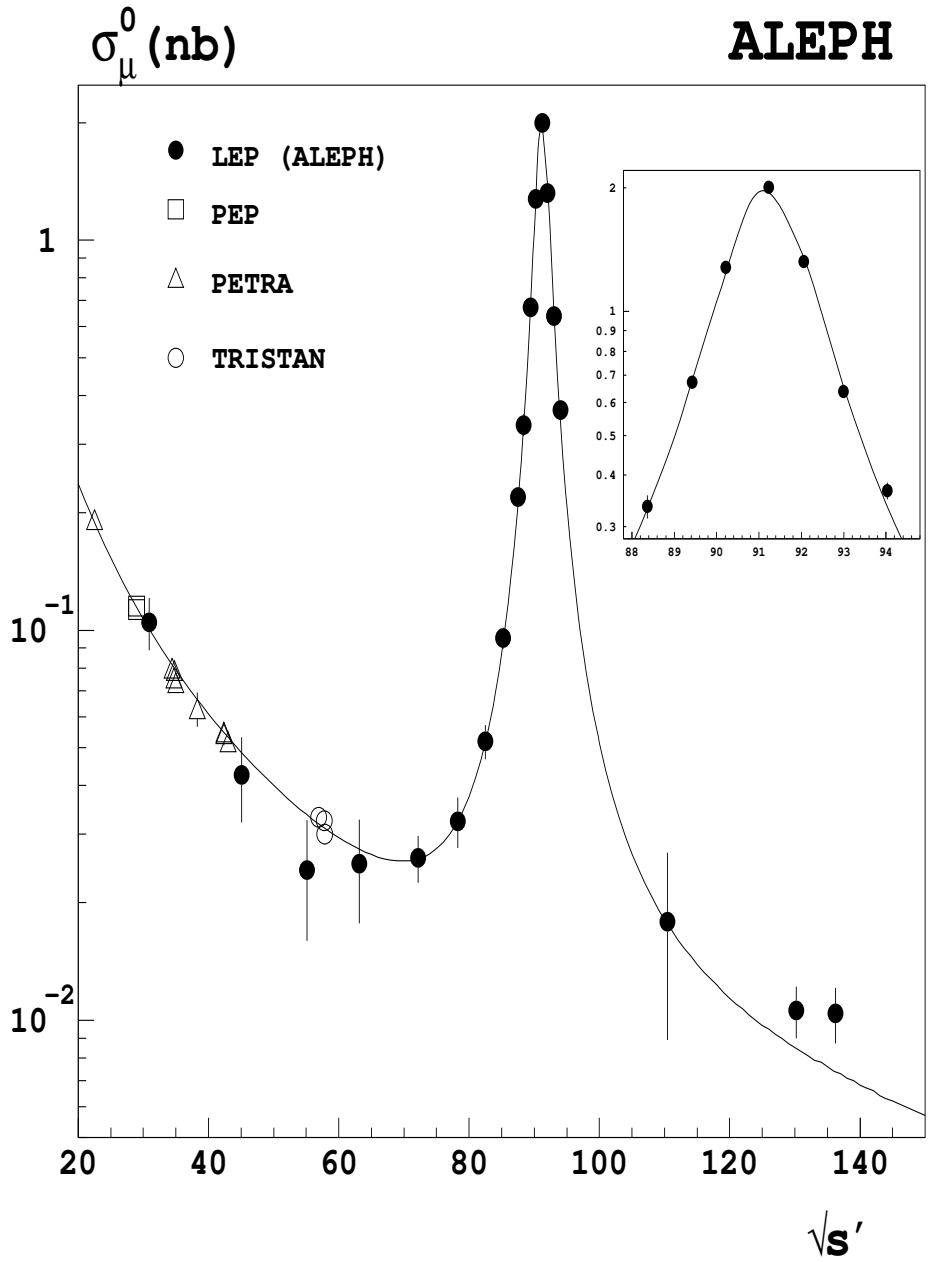,height=32.5cm,%
                    width=18.5cm,%
                   clip=%
}}
 \vspace*{-6.cm}
\begin{center}
\caption{\it
The muon production cross-section over a wide energy range [14]
\label{sigma}
}
\end{center}
\end{figure}

Unfortunately, the width of a particle is not directly measurable.
The $Z$ width may be derived from an analysis of the $Z$ resonance measured at
the accelerators LEP1 and SLC.
There, the most frequent reaction is
\ba
e^+ e^- \to (\gamma, Z) \to {\bar f}f ( + n \gamma)
\ea
The cross-section is shown as a function of the beam energy for a wide energy
range in Figure~\ref{sigma}.

The mass $M_Z$ and width $\Gamma_Z$ 
may be determined from cross-sections obtained in a small region around the
$Z$ peak ($s=4 E_{beam}^2$): 
\ba
|\sqrt{s} - M_Z | < 3~~ \mbox{GeV}
\ea
For energies off the resonance the cross-section falls down rapidly. 

\bigskip

Without radiative corrections, the isolated $Z$ resonance shape may be fitted
with the following ansatz: 
\ba
\sigma_0^{(Z)}(s)&=& \frac{4\pi\alpha^2}{3s} |\chi(s)|^2
\left(a_e^2+ v_e^2\right)\left(a_f^2+v_f^2\right)N_{color}^f
\label{13}
\ea
with a Breit-Wigner shape function
\ba
\chi(s) &=& \frac{G_{\mu} M_Z^2}{\sqrt{2} 2 \pi \alpha} 
   \times   \kappa(s), \qquad \kappa(s) = 
       \frac{s}{s-M_Z^2+iM_Z \Gamma_Z(s)}
\label{breit}
\ea
The natural appearance of an $s$ dependence of the width function in a
perturbative calculation was pointed out by Wetzel (1983)~\cite{wetzel}. 

In Born approximation, the following expression in terms of partial $Z$ widths
is equivalent to~(\ref{13}): 
\ba
\sigma_0^{(Z)}(s)&=& \frac{12\pi}{s} |\kappa(s)|^2 \, 
\frac{ \Gamma_e \, \Gamma_f}{M_Z^2} 
\ea
The cross-section values have to be related to the free
parameters of the theory. 
In the Standard Model these are, e.g., $M_Z, M_H, m_t, \alpha_{strong}$ --
and {\em not}, e.g., $\Gamma_Z$, or the partial widths $\Gamma_f$. 
When radiative corrections are taken into account -- and they have to be --
numerical differences may not be neglected.

\bigskip

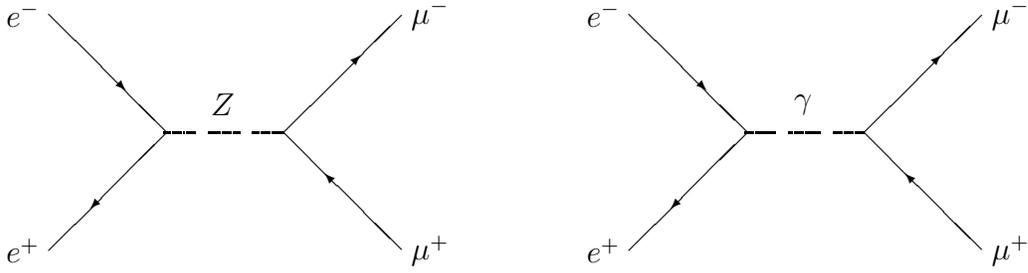
\begin{figure}
{\normalsize \setlength{\unitlength}{1.2mm} \linethickness{0.5pt}
\begin{center}
\begin{picture}(62,51)(0,0)
\put(17.5,37.5){\vector(1,-1){2}}
\put(9.9,44.0){\makebox(0,0)[r]{$e^-$}}
\put(10.9,44.0){\line(1,-1){13.0}} \put(17.5,24.4){\vector(-1,-1){2}}
\put(9.9,17.8){\makebox(0,0)[r]{$e^+$}}
\put(10.9,17.8){\line(1,1){13.0}} \put(30.2,34.0){\makebox(0,0){$Z$}}
\multiput(23.7,30.9)(0.3,0.0){12}{\rule[-0.25pt]{0.5pt}{0.5pt}}
\multiput(28.6,30.9)(0.3,0.0){12}{\rule[-0.25pt]{0.5pt}{0.5pt}}
\multiput(33.5,30.9)(0.3,0.0){12}{\rule[-0.25pt]{0.5pt}{0.5pt}}
\put(43.5,37.5){\vector(1,1){2}}
\put(51.1,44.0){\makebox(0,0)[l]{$\mu^-$}}
\put(37.0,30.9){\line(1,1){13.0}} \put(43.5,24.4){\vector(-1,1){2}}
\put(51.1,17.8){\makebox(0,0)[l]{$\mu^+$}}
\put(37.0,30.9){\line(1,-1){13.0}}
\end{picture} \ 
\begin{picture}(62,51)(0,0)
\put(17.5,37.5){\vector(1,-1){2}}
\put(9.9,44.0){\makebox(0,0)[r]{$e^-$}}
\put(10.9,44.0){\line(1,-1){13.0}} \put(17.5,24.4){\vector(-1,-1){2}}
\put(9.9,17.8){\makebox(0,0)[r]{$e^+$}}
\put(10.9,17.8){\line(1,1){13.0}} \put(30.2,34.0){\makebox(0,0){$\gamma$}}
\multiput(23.7,30.9)(0.3,0.0){12}{\rule[-0.25pt]{0.5pt}{0.5pt}}
\multiput(28.6,30.9)(0.3,0.0){12}{\rule[-0.25pt]{0.5pt}{0.5pt}}
\multiput(33.5,30.9)(0.3,0.0){12}{\rule[-0.25pt]{0.5pt}{0.5pt}}
\put(43.5,37.5){\vector(1,1){2}}
\put(51.1,44.0){\makebox(0,0)[l]{$\mu^-$}}
\put(37.0,30.9){\line(1,1){13.0}} \put(43.5,24.4){\vector(-1,1){2}}
\put(51.1,17.8){\makebox(0,0)[l]{$\mu^+$}}
\put(37.0,30.9){\line(1,-1){13.0}}
\end{picture} \ 
\end{center}
}
\caption{\it
Born contributions to the $Z$ resonance shape
\label{born}
}
\end{figure}

Let me now mention some features of the $Z$ line shape which make its analysis 
complicated.
The first fact is that we want to study a $2 \to 2$ process with intermediate
$Z$, but have also to take into account virtual photon exchange, see
figure~\ref{born}. 

\begin{figure}
{\normalsize \setlength{\unitlength}{0.6mm} \linethickness{0.5pt}
\begin{center}
\begin{picture}(62,77)(0,0)
  \put(31,5){\makebox(0,0)[c] {ISR}} \put(17.5,57.1){\vector(1,0){2}}
  \put(9.9,57.1){\makebox(0,0)[r]{$e^-$}}
  \put(10.9,57.1){\line(1,0){13.0}}
  \multiput(23.7,57.1)(0.3,0.0){12}{\rule[-0.25pt]{0.5pt}{0.5pt}}
  \multiput(28.6,57.1)(0.3,0.0){12}{\rule[-0.25pt]{0.5pt}{0.5pt}}
  \multiput(33.5,57.1)(0.3,0.0){12}{\rule[-0.25pt]{0.5pt}{0.5pt}}
  \put(51.1,70.2){\makebox(0,0)[l]{$\gamma$}}
  \multiput(36.7,57.1)(0.3,0.3){12}{\rule[-0.25pt]{0.5pt}{0.5pt}}
  \multiput(41.6,62.0)(0.3,0.3){12}{\rule[-0.25pt]{0.5pt}{0.5pt}}
  \multiput(46.5,66.9)(0.3,0.3){12}{\rule[-0.25pt]{0.5pt}{0.5pt}}
  \put(24.0,44.0){\vector(0,-1){2}}
  \put(24.0,57.1){\line(0,-1){26.2}} \put(17.5,30.9){\vector(-1,0){2}}
  \put(9.9,30.9){\makebox(0,0)[r]{$e^+$}}
  \put(10.9,30.9){\line(1,0){13.0}}
  \put(30.2,34.0){\makebox(0,0){$A,Z$}}
  \multiput(23.7,30.9)(0.3,0.0){12}{\rule[-0.25pt]{0.5pt}{0.5pt}}
  \multiput(28.6,30.9)(0.3,0.0){12}{\rule[-0.25pt]{0.5pt}{0.5pt}}
  \multiput(33.5,30.9)(0.3,0.0){12}{\rule[-0.25pt]{0.5pt}{0.5pt}}
  \put(43.5,37.5){\vector(1,1){2}}
  \put(51.1,44.0){\makebox(0,0)[l]{$\mu^-$}}
  \put(37.0,30.9){\line(1,1){13.0}} \put(43.5,24.4){\vector(-1,1){2}}
  \put(51.1,17.8){\makebox(0,0)[l]{$\mu^+$}}
  \put(37.0,30.9){\line(1,-1){13.0}}
\end{picture} \ 
\begin{picture}(62,77)(0,0)
  \put(31,5){\makebox(0,0)[c] {ISR}} \put(17.5,57.1){\vector(1,0){2}}
  \put(10.9,57.1){\line(1,0){13.0}}
  \put(30.2,60.2){\makebox(0,0){$A,Z$}}
  \multiput(23.7,57.1)(0.3,0.0){12}{\rule[-0.25pt]{0.5pt}{0.5pt}}
  \multiput(28.6,57.1)(0.3,0.0){12}{\rule[-0.25pt]{0.5pt}{0.5pt}}
  \multiput(33.5,57.1)(0.3,0.0){12}{\rule[-0.25pt]{0.5pt}{0.5pt}}
  \put(43.5,63.6){\vector(1,1){2}}
  \put(51.1,70.2){\makebox(0,0)[l]{$\mu^-$}}
  \put(37.0,57.1){\line(1,1){13.0}} \put(43.5,50.5){\vector(-1,1){2}}
  \put(51.1,44.0){\makebox(0,0)[l]{$\mu^+$}}
  \put(37.0,57.1){\line(1,-1){13.0}} \put(24.0,44.0){\vector(0,-1){2}}
  \put(24.0,57.1){\line(0,-1){26.2}} \put(17.5,30.9){\vector(-1,0){2}}
  \put(10.9,30.9){\line(1,0){13.0}}
  \multiput(23.7,30.9)(0.3,0.0){12}{\rule[-0.25pt]{0.5pt}{0.5pt}}
  \multiput(28.6,30.9)(0.3,0.0){12}{\rule[-0.25pt]{0.5pt}{0.5pt}}
  \multiput(33.5,30.9)(0.3,0.0){12}{\rule[-0.25pt]{0.5pt}{0.5pt}}
  \put(51.1,17.8){\makebox(0,0)[l]{$\gamma$}}
  \multiput(36.7,30.9)(0.3,-0.3){12}{\rule[-0.25pt]{0.5pt}{0.5pt}}
  \multiput(41.6,26.0)(0.3,-0.3){12}{\rule[-0.25pt]{0.5pt}{0.5pt}}
  \multiput(46.5,21.1)(0.3,-0.3){12}{\rule[-0.25pt]{0.5pt}{0.5pt}}
\end{picture} \ 
\begin{picture}(62,77)(0,0)
  \put(31,5){\makebox(0,0)[c] {FSR}} \put(17.5,63.6){\vector(1,-1){2}}
  \put(10.9,70.2){\line(1,-1){13.0}}
  \put(17.5,50.5){\vector(-1,-1){2}}
  \put(10.9,44.0){\line(1,1){13.0}}
  \put(30.2,60.2){\makebox(0,0){$A,Z$}}
  \multiput(23.7,57.1)(0.3,0.0){12}{\rule[-0.25pt]{0.5pt}{0.5pt}}
  \multiput(28.6,57.1)(0.3,0.0){12}{\rule[-0.25pt]{0.5pt}{0.5pt}}
  \multiput(33.5,57.1)(0.3,0.0){12}{\rule[-0.25pt]{0.5pt}{0.5pt}}
  \put(43.5,63.6){\vector(-1,-1){2}}
  \put(51.1,70.2){\makebox(0,0)[l]{$\mu^+$}}
  \put(37.0,57.1){\line(1,1){13.0}} \put(37.0,44.0){\vector(0,-1){2}}
  \put(34.9,44.0){\makebox(0,0)[r]{$\mu^-$}}
  \put(37.0,57.1){\line(0,-1){26.2}} \put(43.5,37.5){\vector(1,1){2}}
  \put(51.1,44.0){\makebox(0,0)[l]{$\mu^-$}}
  \put(37.0,30.9){\line(1,1){13.0}}
  \put(51.1,17.8){\makebox(0,0)[l]{$\gamma$}}
  \multiput(36.7,30.9)(0.3,-0.3){12}{\rule[-0.25pt]{0.5pt}{0.5pt}}
  \multiput(41.6,26.0)(0.3,-0.3){12}{\rule[-0.25pt]{0.5pt}{0.5pt}}
  \multiput(46.5,21.1)(0.3,-0.3){12}{\rule[-0.25pt]{0.5pt}{0.5pt}}
\end{picture} \ 
\begin{picture}(62,77)(0,0)
  \put(31,5){\makebox(0,0)[c] {FSR}} \put(17.5,63.6){\vector(1,-1){2}}
  \put(10.9,70.2){\line(1,-1){13.0}}
  \put(17.5,50.5){\vector(-1,-1){2}}
  \put(10.9,44.0){\line(1,1){13.0}}
  \put(30.2,60.2){\makebox(0,0){$A,Z$}}
  \multiput(23.7,57.1)(0.3,0.0){12}{\rule[-0.25pt]{0.5pt}{0.5pt}}
  \multiput(28.6,57.1)(0.3,0.0){12}{\rule[-0.25pt]{0.5pt}{0.5pt}}
  \multiput(33.5,57.1)(0.3,0.0){12}{\rule[-0.25pt]{0.5pt}{0.5pt}}
  \put(43.5,63.6){\vector(1,1){2}}
  \put(51.1,70.2){\makebox(0,0)[l]{$\mu^-$}}
  \put(37.0,57.1){\line(1,1){13.0}} \put(37.0,44.0){\vector(0,1){2}}
  \put(34.9,44.0){\makebox(0,0)[r]{$\mu^-$}}
  \put(37.0,57.1){\line(0,-1){26.2}}
  \put(43.5,37.5){\vector(-1,-1){2}}
  \put(51.1,44.0){\makebox(0,0)[l]{$\mu^+$}}
  \put(37.0,30.9){\line(1,1){13.0}}
  \put(51.1,17.8){\makebox(0,0)[l]{$\gamma$}}
  \multiput(36.7,30.9)(0.3,-0.3){12}{\rule[-0.25pt]{0.5pt}{0.5pt}}
  \multiput(41.6,26.0)(0.3,-0.3){12}{\rule[-0.25pt]{0.5pt}{0.5pt}}
  \multiput(46.5,21.1)(0.3,-0.3){12}{\rule[-0.25pt]{0.5pt}{0.5pt}}
\end{picture} \ 
\end{center}
}
\caption{\it
QED corrections to the $Z$ resonance shape
\label{qed}
}
\end{figure}
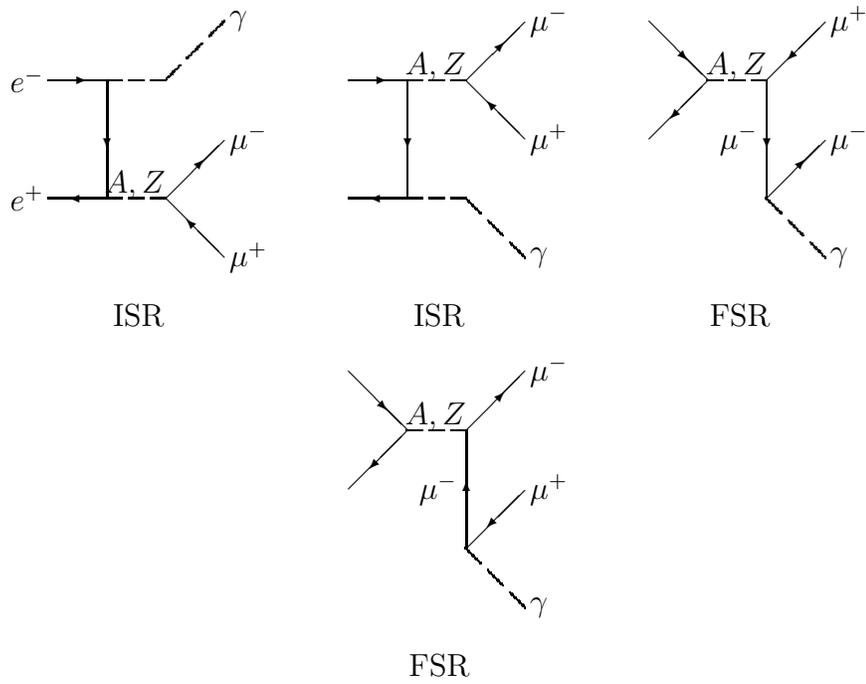

In addition, there are huge $2 \to 3,4,\ldots$ contributions due to initial
state 
radiation (ISR) and final state radiation (FSR), see figure~\ref{qed}.

Further, many virtual corrections are not shown here but have to be included:
\begin{itemize}
\item vertex insertions
\item self-energy insertions
\item box diagrams
\item for quarks: final state QCD corrections
\end{itemize}

In the next section, I will indicate the proper handling of the $Z$ resonance
as it is practized by the LEP experimental groups ALEPH, DELPHI, L3, OPAL and
the SLD group at SLC.
Further details may be found in~\cite{blondel,analysis} and, of course, in the
articles of these collaborations. 
\section{Photonic Corrections to the $Z$ Line Shape
}
Photonic corrections influence position and shape of the $Z$ resonance heavily.
An analysis of data without proper treatment of them leads to results which are
numerically simply wrong. 
Fortunately, they may be taken into account in a generic way by the following
convolution formula: 
\ba
\sigma(s)
 &=&\int d \left(s'/s\right) \sigma_0(s')\, \rho\left( s'/s \right)
~+~\int d \left(s'/s\right) \sigma_0^{int}(s,s')
\, \rho^{int} \left( s'/s \right)
\label{sigqed2}
\ea
with
\begin{itemize}
\item
$\rho(s'/s)$ -- a radiator 
describing initial and final state radiation of photons,
including leading higher order effects and soft photon exponentiation;
\item
$\rho^{int}(s'/s)$ -- taking into account the 
initial-final state interference effects which
are comparatively small (a few per mille);
\item
$\sigma_0(s')$ -- the basic scattering cross-section, which is the
object of investigation;
\item
$\sigma_0^{int}(s,s')$ -- a similar function, but often negligible since near
the resonance peak numerically suppressed.
\end{itemize}
If $\rho$ is known and $\rho^{int}$ is also known and, more important, small,
one may try to unfold the basic cross-section $\sigma_0$
from experimental data.

The dominant part of the photonic corrections is due to ISR (initial state
radiation):
\ba
\rho(s'/s) &=& \beta (1-s'/s)^{\beta-1}\delta^{soft+virtual}+\delta^{hard}
\label{rho}
\ea
where~\cite{analysis}
\ba
\beta &=&
 \left(2\alpha/\pi\right) \left[\ln\left(s/m_e^2\right)-1\right]
\\
\delta^{soft+virtual} &=&
1
+ \left(\alpha/(2\pi)\right) \left[3\ln\left(s/m_e^2\right) +2 \pi^2 /3
-4 \right] 
+{\cal O} \left(\alpha^2\right)
\\
\delta^{hard} &=& -\left(\alpha/\pi\right) (1+s'/s)
\left[\ln\left(s/m_e^2\right)-1\right] 
+{\cal O} \left(\alpha^2\right)
\ea
Near the resonance peak ISR leads to huge corrections
of shape, position, and height of the peak and cannot be neglected.
This means that any serious physical analysis is not only faced by higher order
corrections but also by a substantial admixture of $2 \to 3$ (and higher order)
processes which may not be experimentally extracted from the $2 \to 2 $ process
under investigation.     
\section{Approaching a Reasonable Formula for $\sigma_0$
}
In this section, I discuss the sensitivity of the determination of $M_Z$ on the
theoretical ansatz\footnote{
More details may be found in~\cite{marck90}.}.
At first glance, one may expect that the peak of the $Z$ resonance is at
$\sqrt{s_{\max}} = M_Z$.
This would reduce the determination of $M_Z$ to a search of the peak location. 
This intuitive picture is modified by several effects.
As already mentioned, photonic initial state radiation is rather influential.
One may estimate the resulting shift of the $Z$ resonance peak
from~(\ref{rho}):
\ba
\sqrt{s_{\max}} - M_Z &=& 
\delta_{QED} =
\frac{\pi}{8} \beta 
\Gamma_Z
+ \,\mbox{small corr's.}
\approx 106 \,\mbox{MeV}
\label{shift00}
\ea
A simple and reasonable ansatz for $\sigma_0^{(Z)}$ is a pure Breit-Wigner
function 
\ba 
\sigma_0^{(Z)}(s) \sim
\frac {M_Z^2\cdot  R } {\left|s-M_Z^2 +i M_Z \Gamma_Z\right|^2}
\label{breit2}
\ea
It may be shown (and will be made plausible in the next two sections) that the
following  ansatz is more realistic\footnote{This or similar formulae
have been proposed in~\cite{borrelli,jegerl,stuart262,riemann}.}:  
\ba
\sigma_0(s) =
\frac{4}{3} \pi \alpha^2
\left[ \frac{r^{\gamma}}{s} +
\frac {s\cdot R + (s - M_Z^2)\cdot J} 
{\left|s-M_Z^2 + i s \Gamma_Z/M_Z\right|^2}
\right]
\label{sigqed3}
\ea
This line shape is characterized by five parameters:
\begin{itemize}
\item $r^{\gamma} \sim \alpha_{em}^2(M_Z^2)/\alpha_{em}^2$ -- this $\gamma$
exchange term may be assumed to be known 
\item  $M_Z,~~ \Gamma_Z$
\item $R$ -- measure of the $Z$ peak height
\item $J$ -- measure of the $\gamma Z$ interference
\end{itemize}
Besides~(21),
from the replacements
\ba
M_Z^2\cdot  R \to s \cdot  R, ~~~i M_Z \Gamma_Z \to i s \Gamma_Z / M_Z
\ea
additional shifts arise:
\ba
\sqrt{s_{\max}} - M_Z &=& 
\delta_{QED} \oplus 
\frac{1}{4} \frac{\Gamma_Z^2}{M_Z} 
\ominus
\frac{1}{2}\frac{\Gamma_Z^2}{M_Z} \sim \left( 90+ 17 - 34\right)~~\mbox{MeV} 
\ea
Additionally, there is the effect of the $\gamma Z$ interference $J$: 
\ba
\sqrt{s_{\max}} - M_Z 
&=& 
\delta_{QED} \oplus 
\frac{1}{4} \frac{\Gamma_Z^2}{M_Z} \left(1+\frac{J}{R}\right)
\ominus
\frac{1}{2}\frac{\Gamma_Z^2}{M_Z}
\nl
&\sim& \left[90 + 17 \times \left(1+\frac{J}{R}\right) -34\right]
~~\mbox{MeV} 
\ea
If one wants to take into account the $J$, a model for its
prediction is needed.
Neglecting this interference (by setting $J$=0) leads to an erroneous
systematic shift of the $Z$ mass of 17~MeV$\otimes(J/R)$.

The value for hadron production in the Standard Model is,
e.g.,~\cite{lepewwg96}: 
\ba
\frac{J}{R}\otimes 17~~\mbox{MeV} = \frac{0.22}{2.97}\otimes 17~~\mbox{MeV}
= 0.07\otimes 17~~\mbox{MeV} = 1.26 ~~\mbox{MeV}
\ea
\section{$Z$ Boson Parameters (I): 
\\
\mbox{The Standard Model Approach}
\label{I}
}
A realistic scan of the $Z$ line shape may be performed with the following
ansatz derived from the Standard Model, including higher order radiative
corrections~\cite{analysis}: 
\ba
\sigma_0(s) &=& \sigma_0^{(Z)}(s) + \sigma_0^{(\gamma Z)}(s) +
\sigma_0^{(\gamma)}(s)
\label{sigsm}
\ea
The dominating part is
\ba
\sigma_0^{(Z)}(s)&=& \frac{4\pi\alpha^2}{3s} |\chi(s)|^2
|\rho_{ef}^{eff}|^2 
\left(
\frac{1}{16} + \frac{1}{4} |v_e^{eff}|^2 +
\frac{1}{4} |v_f^{eff}|^2 + |v_{ef}^{eff}|^2
\right)
\label{sigsm1}
\ea
with an $s$-dependent width function in the Breit-Wigner shape~(\ref{breit}).
The width is obtained in perturbation theory by summing an infinite Dyson
series of self-energy insertions to the $Z$ boson propagator.
In order to prevent gauge violation (in the given order of perturbation
theory), one has to add up a minimal set of Feynman diagrams that is
necessary for the compensation of gauge dependences.
The decay width of the $Z$ boson in lowest order is given by the
imaginary parts of fermion loops in the one loop self-energy.
For single $Z$ boson production we may use\footnote{
For the by far more complex case of off-shell gauge boson pair
production, the fermion-loop scheme~\cite{argyres,beenakker} solves
the gauge problem satisfactorily.  
}: 
\ba
\Gamma_Z(s) &=& \frac{s}{M_Z^2}\Gamma_Z
\label{gamsm}
\ea
The simple $s$ dependence is due to the smallness of the fermion masses
allowing for the neglect of threshold effects.
A complete two-loop calculation would also modify this. 
The virtual corrections are contained in the width $\Gamma_Z$ and in four
complex-valued form factors $\rho_Z^{eff}$, $v_e^{eff}$, $v_f^{eff}$,
$v_{ef}^{eff}$ which depend on beam energy and scattering angle.

To a good approximation, (\ref{sigsm}) agrees with~(\ref{sigqed3}).
The cross-section values have to be related to the free
parameters of the theory. 
A recent determination is~\cite{l396}: 
\ba
M_Z &=& 91.188 \pm 0.002~~ \mbox{GeV}
\\
\alpha_{strong}(M_Z^2) &=& 0.126 \pm 0.007 \pm 0.002~\mbox{(Higgs)}
\\
m_t &=& 189 \pm 15~\mbox{(exp.)} \pm 16~\mbox{(Higgs)} ~~~\mbox{GeV}.
\ea
The Higgs mass is varied from 60 to 1000 GeV with central values for 300 GeV.
Quite similar values have been quoted in~\cite{blondel}. 
For comparison I quote also the direct measurements of
$m_t$ by CDF, $m_t = 175.6 \pm 4.4 \pm 4.8$ GeV, and D0
collaborations, $m_t = 169 \pm 8 \pm 8$ GeV~\cite{cdfd0}.

A step towards a model-independent $Z$ resonance analysis is the
determination of some characteristic line shape parameters from table~6
of~\cite{blondel} 
(with indicated relative errors $\delta$):
\ba
M_Z &=&  91.1863 \pm 0.0020~~ \mbox{GeV}~~~~(\delta=0.0022~\%)
\\
\Gamma_Z &=& 2.4946 \pm 0.0027~~ \mbox{GeV}~~~~(\delta=0.11~\%)
\\
\sigma_0^{had} &=& 41.508 \pm 0.056~~{\mbox{nb}}~~~~(\delta=0.13~\%)
\\
R_l = \frac{\sigma_0^{had}}{\sigma_0^{lept}} &=& 20.778 \pm
0.029~~~~(\delta=0.14~\%) 
\\
A_{FB,0}^{lept} &=& 0.0174 \pm 0.0010
\ea
Here, $M_Z, \Gamma_Z, \sigma_0^{had}$ are determined mainly from
$\sigma^{had}(s)$, while $R_l$ and $A_{FB}$ from $\sigma^{lept}(s)$:
\begin{itemize}
\item $\sigma_0^{had(lep)}$ -- hadronic (leptonic) peak cross-section
\item $A_{FB,0}^{lept}$ -- forward-backward asymmetry at the peak
\end{itemize}
These parameters are considered to be primary parameters in contrast
to derived ones, e.g. the effective
leptonic weak neutral current couplings of leptons (table~8 of~\cite{blondel}):
\ba
v_l &=& - 0.03776 \pm 0.0062
\\
a_l &=& - 0.50108 \pm 0.00034
\ea
or the effective
weak mixing angle (tables~1,4 of~\cite{blondel}): 
\ba
\sin^{2}\vartheta_{W}^{eff} 
\equiv \frac{1}{4}\left( 1-\frac{v_l}{a_l}\right)
&=& 0.23165 \pm 0.00024
\label{sw1}
\ea
The introduction of effective weak neutral couplings and the effective weak
mixing angle comes back close to the language of the Standard Model.
\section{$Z$ Boson Parameters (II):  
\\ 
\mbox{The {S-Matrix} Approach}
}
All the above results fit nicely with each other and strengthen the Standard
Model's credit.
Nevertheless, one may ask for an approach being independent of it.
A tool with minimal assumptions is S-matrix
theory~\cite{olive,bohm,martin}. 
The first application of S-matrix theory to the $Z$ resonance is due to A.
Martin (1985)~\cite{martin} who studied the toponium-$Z$ interference pattern
assuming their masses to be of similar size.
In 1991, R. Stuart proposed to consider the scattering matrix element for
the process $e^+e^- \to Z \to f {\bar f}$ as a Laurent series with the $Z$
boson as resonance~\cite{stuart262}.
This allowed him to collect gauge invariant pieces of the cross-section
in perturbation theory~\cite{stuart272} and to derive a simple
cross-section formula similar 
to~(\ref{breit2}), but with small perturbations.  
For an application to experimental data, a number of modifications have been
added~\cite{riemann,riemann92} and the necessary software has been
created~\cite{riemann95}: 
consider the cross-section as an incoherent sum of four helicity scatterings;
treat the photonic corrections properly, especially those due to initial state
radiation; 
treat in the same manner as the total cross-section also asymmetries;
try to include into the formula the fact that there is also photon exchange,
i.e. that in reality one has the co-existence of two resonances.  
The first fit to LEP~1 data was performed in~\cite{riemann}.
\bigskip
     
Consider 
four independent helicity amplitudes in the case of massless fermions $f$:
\ba
\label{eqn:mat0}
{\cal M}^{fi}(s) = \frac{\Rgf}{s} + \frac{\RZfi}{s-s_Z} +
\sum_{n=0}^\infty \frac{\Ffin}{\ovMZ^2} \left(\frac{s-s_Z}{\ovMZ}\right)^n 
,~~\  i=0,\ldots,3
\ea
Without the first (photon) term, they are Laurent series.
The position of the \Zo\ pole in the complex $s$ plane is given by 
$s_{\Zo}$:
\ba
\label{szb}
s_Z = \ovMZ^2 - i \ovMZ \ovGZ
\ea
The \Rgf\ and \RZfi\ are complex constants characterizing the photon and the
\Zo\ boson, respectively. 
For practical purposes one may truncate the series:
\ba
\label{eqn:mat}
{\cal M}^{fi}(s) 
&=& \frac{\Rgf}{s} + \frac{\RZfi}{s-s_Z} +
\frac{F_0^{fi}}{\ovMZ^2}
\approx
\frac{\Rgf}{s} + \frac{\RZfi}{s-s_Z} 
\ea
There are four residua \RZfi:
\ba
\renewcommand{\arraystretch}{1.2}
\label{eqn:mi_he14}
\begin{array}{lcl}
\RZfn & = & R_Z (e^-_Le^+_R \longrightarrow f^-_L f^+_R) \\
\RZfe & = & R_Z (e^-_Le^+_R \longrightarrow f^-_R f^+_L) \\
\RZfz & = & R_Z (e^-_Re^+_L \longrightarrow f^-_R f^+_L) \\
\RZfd & = & R_Z (e^-_Re^+_L \longrightarrow f^-_L f^+_R)
\end{array}
\renewcommand{\arraystretch}{1.}
\ea
The amplitudes
${\cal M}^{fi}(s)$ give rise to four cross-sections $\sigma_i$:
\ba
\renewcommand{\arraystretch}{1.2}
\label{eqn:xs14}
\begin{array}{lclcl}
&&\sigma_{T}^0(s)      & = & +~  \sigma_0 + \sigma_1 + \sigma_2 + \sigma_3
  \\
\sigma_{\mbox{\scriptsize \it lr-pol}}^0(s) & = &
\frac{4}{3}\sigma_{FB}^0(s)     
& = & +~  \sigma_0 - \sigma_1 + \sigma_2 - \sigma_3
 \\
\frac{4}{3}\sigma_{\mbox{\scriptsize \it FB-lr}}^0(s)& = &
\sigma_{pol}^0(s)    & = & -~  \sigma_0 + \sigma_1 + \sigma_2 - \sigma_3
 \\
\sigma_{lr}^0(s)& = &
\frac{4}{3}\sigma_{\mbox{\scriptsize \it FB-pol}}^0(s) & = & -~  \sigma_0 -
\sigma_1 + \sigma_2 + \sigma_3
\end{array}
\renewcommand{\arraystretch}{1.}
\ea
Here, it is

$\sigma_{T}^0$ -- the total cross-section, 

$\sigma_{FB}^0$ -- numerator of the forward-backward asymmetry, 

$\sigma_{pol}^0$ -- numerator of the final state polarization,

$\sigma_{\mbox{\scriptsize \it FB-pol}}^0$ -- that of the forward-backward
asymmetry of the 
final state polarization etc.
\\
All these cross-sections may be parameterized by the following master
formula: 
\ba
\renewcommand{\arraystretch}{2.2}
\label{eqn:smxs}
\sigma_A^0(s)
&=&
\displaystyle{\frac{4}{3} \pi \alpha^2
\left[ \frac{r^{\gamma f}_A}{s} +
\frac {s r^f_A + (s - \ovMZ^2) j^f_A} {(s-\ovMZ^2)^2 + \ovMZ^2 \ovGZ^2} 
+ \frac{r_A^{f0}}{\ovMZ^2}\right]} + \ldots
\nl
&\approx&
\displaystyle{\frac{4}{3} \pi \alpha^2
\left[ \frac{r^{\gamma f}_A}{s} +
\frac {s r^f_A + (s - \ovMZ^2) j^f_A} {(s-\ovMZ^2)^2 + \ovMZ^2 \ovGZ^2}
\right]}
, \hspace{.5cm} A = \mbox{\it T,} \ldots, 
\renewcommand{\arraystretch}{1.}
\ea
where the $\sigma_{\mbox{\scriptsize \it FB-\ldots}}^0(s)$ again get
an additional factor 3/4.
\\
Thus we have re-derived~(\ref{sigqed3}) with one modification:
the $Z$ width function is treated as constant here.
Now, the parameters $r,j$ are related to the residua of the pole terms. 
The $r^{\gamma f}_A$ is the photon exchange term:
\ba
\label{qedr}
r^{\gamma f}_A =
\displaystyle{\frac{1}{4}  c_f \sum_{i=0}^3 \{\pm 1\}
\left| \Rgf \right|^2 } R_{QCD}^A
\ea
It is known from QED for the total cross-section ($A=T$) and vanishes for all
asymmetric cross-sections. 
Further, $c_f=1,3$ for leptons and quarks, respectively.
QCD corrections for quarks are taken into account by the factor $R_{QCD}$.
The \Zo\ exchange residuum $r^f_A$ and the \gam\Zo\ interferences
$j^f_A$ are:
\ba
\label{eqn:rrjj}
\renewcommand{\arraystretch}{2.}
\begin{array}{lll}
r^f_A & = & \displaystyle{c_f \left\{\frac{1}{4}\sum_{i=0}^3 \{\pm 1\}
\left| \RZfi \right|^2 + 2 \frac{\ovGZ}{\ovMZ} \imag C^f_A \right\}} R_{QCD}^A
\\ 
j^f_A & = & c_f\left\{2 \real C^f_A - 2 \displaystyle{\frac{\ovGZ}{\ovMZ}}
\imag C^f_A\right\} R_{QCD}^A
\\
C^f_A & = & (R^f_\gamma)^* \left(\displaystyle{\frac{1}{4}\sum_{i=0}^3
\{\pm 1\} \RZfi}\right) 
\end{array}
\renewcommand{\arraystretch}{1.}
\ea
The factors $\{\pm 1\}$ in~(\ref{qedr}) and~(\ref{eqn:rrjj})
indicate that the signs of $\left| R_{\gamma}^f \right|^2$, 
$\left| \RZfi \right| ^2$,
and of \RZfi\ correspond to the signs of $\sigma_i$ in~(\ref{eqn:xs14}).
\subsection{Asymmetries}
Without QED corrections, asymmetries are defined by:
\ba
\label{eqn:mi_asy}
{\cal A}_A^0(s) = \frac{\sigma_A^0(s)}{\sigma_{T}^0(s)},~~~ A \neq T
\ea

\begin{figure}[thbp]
\begin{center}
\mbox{
        \epsfig{file=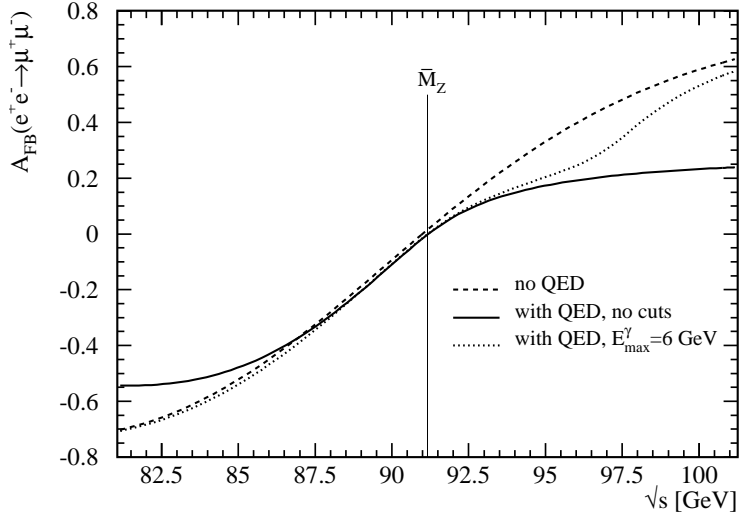,
                   width=10cm,%
                   clip=%
             }
     }

\end{center}
\caption[foo]{
The forward-backward asymmetry for the process $e^+ e^- \rightarrow \mu^+
\mu^-$ near the \Zo\ peak~[26].
\label{asy}
}
\end{figure}
They take a simple form around the \Zo\
resonance.
For applications at LEP~1, they may be characterized by only two
parameters~\cite{riemann92}:  
\bq
{\cal A}_A^0(s) = A_0^A + A_1^A \left(\frac{s}{\ovMZ^2} - 1 \right) +
             A_2^A \left(\frac{s}{\ovMZ^2} - 1 \right)^2 + \ldots
\approx
A_0^A + A_1^A \left(\frac{s}{\ovMZ^2} - 1 \right)
\label{e1}
\eq
The higher order terms may be neglected since
$(s/\ovMZ^2-1)^2 = 
\sigma^2 < 2 \times 10^{-4}$.
The first coefficients are:
\ba
\label{eqn:a0}
A_0^A &=&
\frac{r_A^f}{r_T^f + \gamma^2 r^{\gam f}_T} \approx \frac{r_A^f}
{r_T^f}
\\
\label{eqn:a1}
A_1^A &=&
\left[ \frac{j^f_A}{r_A^f} - \frac{j_T^f - 2\gam^2 r^{\gam f}_T}
{r_T^f+ \gam^2 r^{\gam f}_T} \right] A_0^A
\approx
\left[ \frac{j^f_A}{r_A^f} -  \frac{j_T^f}{r_T^f} \right] A_0^A
\label{e15}
\ea
Here, the $r_A^{0f}$ is neglected in both $A_0$ and $A_1$. 
Further, the definition
$\gamma^2 = \ovGZ^2 / \ovMZ^2 \approx 0.75 \times 10^{-3}$ is used.
The non-vanishing of the slope of the asymmetry shape is due to the $\gamma Z$
interference. 
From figure~\ref{asy} one may see that the linear rise is damped right of the
peak.
This is due to amplified QED corrections to the pure $Z$ exchange
cross-sections versus non-amplified QED corrections to the $\gamma Z$
interferences~\cite{riemann92}.  

\subsection{Numerical Results in the S-matrix Approach}
The first fit with the S-matrix approach to experimental data has been
performed in 1991~\cite{riemann}.
The first experimental analysis of a LEP collaboration was due to
L3 in 1993~\cite{l3zz}. 
Further systematic studies may be found
in~\cite{kirsch92}.     
They helped to determine the appropriate number and location of energy points
for a $Z$ line shape scan.
Recent experimental studies are
e.g.~\cite{martinez95,lepewwg95,lepewwg96,opal9602,l396}. 
Typically, results as in table~\ref{table1} are obtained from the LEP1 and
LEP1.5 $Z$ line shape scans which were performed mainly in 1993 and
1995 (from table~6 of~\cite{lepewwg96}\footnote{Note that the table
  shows values of the on shell mass $M_Z$ which were derived 
after the fit of the complex pole mass $\ovMZ$ as explained in
section~\ref{comm1}.  
}).
The biggest error correlations are shown in table~2 (from table~7
of~\cite{lepewwg96}). 
We see here an essential difference to Standard Model fits which assume {\em
fixed} relations among many of these parameters. 
They rely thus on stronger theoretical assumptions. 

{\em From the strong correlations in the S-matrix fit together with the excellent
agreement of the central values of fitted parameters in both fit scenarios one
may conclude that the two scenarios are highly compatible with each other.}


\begin{table}[htbp]\centering
\begin{tabular}{|c|r@{$\pm$}l|r@{.}l|}
\hline
Parameter & \multicolumn{2}{c|}{S-matrix fit} & 
            \multicolumn{2}{c|}{SM Prediction}
\\
\hline
\hline
$M_Z$ [GeV]      &  91.1965&0.0048  & \multicolumn{2}{c|}{--} 
\\
$\Gamma_Z$ [GeV] &   2.4941&0.0033  &  2&4973
\\
\hline
$r_T^{had}$      &  2.9644   & 0.0083   & 2&9681 
\\
$j_T^{had}$     &   0.22    & 0.25   &  0&22
\\
\hline
$r_T^{lept}$     &   0.14245 & 0.00044&  0&14268
\\
$j_T^{lept}$     &   0.020   & 0.017  &  0&004
\\
\hline
$r_{FB}^{lept}$  &   0.00315 & 0.00022&  0&00271
\\
$j_{FB}^{lept}$  &   0.793   & 0.016 & 0&799
\\  \hline
\end{tabular}
\caption
{\it
Results from a combined LEP1 line shape fit
\label{table1}
}
\end{table}


\begin{table}[thbp]\centering
\begin{tabular}{|r@{--}l|r@{.}l|}
\hline
\multicolumn{2}{|c|}{Correlation} & \multicolumn{2}{c|}{Value}
\\
\hline
\hline
$M_Z $ & $ j_T^{had}$  & --0&89
\\
$M_Z $ & $ j_T^{lept}$ & --0&62
\\
$\Gamma_Z $ & $ r_T^{had}$ & 0&77
\\
$\Gamma_Z $ & $ r_T^{lept}$ & 0&69
\\
$r_T^{had} $ & $ r_T^{lept}$ & 0&86
\\
$j_T^{had} $ & $ j_T^{lept}$ & 0&62
\\  \hline
\end{tabular}
\caption
{\it
Biggest correlations in the S-matrix fit
\label{table2}
}
\end{table}



Including into the analysis cross-sections measured at other energies may
improve substantially e.g. the resolution of $M_Z$ and $j_T$ which are highly
correlated (for a combination with data from the 
TOPAZ collaboration at KEK
 with $\sqrt{s} \sim$ 55 GeV as shown in figure~1 see
reference~\cite{l3note96}; data from LEP~1.5 
 with $\sqrt{s} \sim$ 135 GeV have been included already). 
\section{ 
Defining the $Z$ Boson Mass
\label{comm1}
}
The complex $Z$ pole definition 
in~(\ref{eqn:mat0})  
with a constant width is
natural in the S-matrix ansatz.
It leads to different numerical values compared to 
the usual Standard Model, on mass shell approach as used
in~(\ref{sigsm})--(\ref{gamsm}).
The following discussion of this difference follows closely appendix~D
of~\cite{beenakker} where more details may be found.

In the perturbative approach, the {\em complex pole} $\mu_Z$ of the propagator
is defined as follows: 
\ba
\mu_Z - \mu_Z^0 + \Sigma_Z^0(\mu_Z,\mu_Z^0) &=& 0
\\
\mu_Z &=&  \ovMZ^2 - i \ovMZ \ovGZ
\ea
The bare $Z$ boson mass is denoted by $\mu_Z^0$ and $\Sigma_Z^0$ is the bare
self-energy.
The perturbative solution of the above equations is:
\ba
\ovMZ^2 &=& \mu_Z^0 - \Re e \Sigma_Z^0 (\ovMZ^2) - 
\left[\Im m \Sigma_Z^0 (\ovMZ^2)\right] 
\left[\Im m \Sigma_Z^{0\prime}(\ovMZ^2)\right] 
+ \ldots
\\
\ovMZ \ovGZ &=& \Im m \Sigma_Z^0 (\ovMZ^2) 
\Biggl\{
1 - \Re e \Sigma_Z^{0\prime} (\ovMZ^2) + \left[ \Re e
\Sigma_Z^{0\prime} (\ovMZ^2) 
\right]^2 
\nl &&
-~ \frac{1}{2}  
\left[\Im m \Sigma_Z^{0}(\ovMZ^2)\right] 
\left[\Im m \Sigma_Z^{0\prime\prime}(\ovMZ^2)\right] + \ldots 
\Biggr\}
\ea
The {\em on shell mass and width} are defined as:
\ba
M_Z^2 &=& \mu_Z^0 - \Re e \Sigma_Z^0 (M_Z^2)
\\
M_Z \Gamma_Z &=& \frac{\Im m \Sigma_Z^0 (M_Z^2)}
{1+\Re e \Sigma_Z^{0\prime}(M_Z^2)}
\ea
One may relate the two definitions and see that they differ by two-loop
and higher order corrections.
This has been discussed first in 1986~\cite{consoli}. 
The expected experimental accuracy was about 10 MeV~\cite{altarelli}
at that time.
Since the authors of~\cite{consoli} restricted themselves to the one
loop order, they failed to observe the numerical significance of the
difference in the definitions of about 35 MeV. 
There are also bosonic corrections in one-loop approximation. 
A systematic Dyson summation of bosonic self-energy corrections may be
attempted without violating Ward identities in the context of the
background field method~\cite{denner}.
For references to the application of so-called pinch techniques see also there.
Since the on shell mass and the complex mass definitions are uniquely
related order by order in perturbation theory, it is fair to say that
either both or none of them has a gauge invariance problem, provided
it is used properly.
One may argue that a mass definition should be related to a structure
like Const/($s-s_0$), with $s_0$ being a constant, but as long as
relations are unique to another definition, there is some freedom of
choice.
Of course, the (perturbative) complex mass definition is conceptually
closest to what one has in the S-matrix theory.

Around $s=M_Z^2$ the $Z$ decays only into light fermions and it is
\ba
\Im m \Sigma_Z^0 (s) = s ~ \Im m \Sigma_Z^{0\prime} (s) = s ~
\frac{\Gamma_Z}{M_Z} 
\ea
and thus
\ba
\ovMZ^2 &=& M_Z^2 - \Gamma_Z^2 + \ldots
\\
\frac{\ovGZ}{\ovMZ} &=& \frac{\Gamma_Z}{M_Z} 
\ea
The resulting numerical differences may well be approximated by the
following 
relations:
\ba
\label{mmggzz}
\renewcommand{\arraystretch}{1.2}
\begin{array}{lllllcr} 
\ovMZ & = & M_Z - \frac{\displaystyle \Gamma_Z^2}{\displaystyle 2M_Z}
&\approx& M_Z&-&34~{\mbox{MeV}} 
\\
{\overline \Gamma}_{\Zo} & = & \Gamma_Z -  \frac{\displaystyle
\Gamma_Z^3}{\displaystyle 2M_Z^2} 
&\approx& \Gamma_Z& - &1~\mbox{MeV}. 
\end{array}
\ea
\renewcommand{\arraystretch}{1.}
Equations~(\ref{mmggzz}) and the numerical values of the shifts were
derived in 1988 in~\cite{leike}, where also the  
$\gamma Z$ mixing was taken into account, and repeatedly
discussed later (see
e.g.~\cite{burgers,jegerTasi,stuart262,willenbrock,lopez0}).    
The $Z$ resonance peak shift due to the difference of the two
treatments of the $Z$ boson self energy 
was numerically observed independently in~\cite{BBHvN} and
in~\cite{leike}.
Both papers did not point out the significance of the complex pole
mass definition, although it became obvious immediately after and was
frequently discussed during the 1989 LEP~1 workshop organized by CERN.  
Similar derivations to the above may be found
in~\cite{consoli,jegerTasi,stuart262,sirlin}.  
Gauge problems have been studied also in~\cite{HV}. 

\bigskip

The observation of the sensitive dependence of numerical mass values on the
definitions chosen in the Breit-Wigner shape function was made for
hadron resonances by Gounaris and Sakurai (1968) \cite{gounaris}.
See also the recent studies of hadron resonances reviewed
in~\cite{lopez}.  
\section{The $Z$ Resonance and the Photon}
Being strict, one may develop the S-matrix into a Laurent series as
a function of $s$ around {\em one} resonance only.
Otherwise the coefficients are not uniquely determined.  
This has been stressed recently~\cite{stuart2}. 

In order to be rigorous, one has to replace the ansatz~(\ref{eqn:mat0}) 
by one with $\Rgf = 0$.
The essential physical consequence is that the photonic cross-section becomes
part of the 
background: 
\ba
\frac{r^{\gamma}}{s} \to \frac{r^{\gamma}}{M_Z^2}
\ea
What was understood so
far as $\gamma Z$ interference $J$ becomes a result of the
background-$Z$ interference. 

I, personally, dislike this approach. The photon exists and we know how to
describe it. 
So, I would prefer to see it treated as known input to an experimental analysis
of the $Z$.
But I agree that a detailed study of the resulting numerical differences
between the rigorous S-matrix ansatz and that used by the LEP community at
present could be of some interest.
Perhaps it is worth to be mentioned that a consistent quantum
mechanical description of two-resonance systems is possible; see
e.g. chapter~XX in~\cite{bohm}.   
\section{Summary}
Two different approaches to a numerical
analysis of the $Z$ boson line shape have been presented -- the Standard Model
of electroweak interactions and the S-matrix approach.
The S-matrix approach allows to treat the $\gamma Z$ interference as an
independent quantity, which enlarges the error for $M_Z$.
Two different mass definitions may be used.
Both agree in the numerical determination of the $Z$ mass when the
substantial difference in the mass definitions is taken into account.
In the Standard Model the $Z$ width is a derived quantity;
the S-matrix approach allows a direct fit.
Again, the two approaches agree numerically.
The S-matrix approach shows that the $Z$ line shape may be described
by 4 independent parameters (per channel) -- if QED is assumed to be a
known 
phenomenon.
Asymmetries may also be described by the   S-matrix approach.
They depend on two parameters (per channel).
Their variation with $s$ near the peak is due to the $\gamma Z$
interference. 
The S-matrix approach allows the combination of data from quite
distinct kinematic regions.
Finally, one should mention that the use of effective coupling
constants leads 
to similarly reasonable numerical results.
\section*{Acknowledgements}
I would like to thank the organizer of the Symposium on Semigroups and
Resonances, Arno B\"ohm, for the kind invitation to the conference
and to Wim Beenakker, Fred Jegerlehner, and Robin Stuart for careful
reading the manuscript and suggestions. 

\end{document}